\documentstyle[preprint,aps,prd]{revtex}

\begin{document}

\draft
\title{Geodesics, gravitons and the gauge fixing problem}

\author{Diego A. R. Dalvit$^{1}$\thanks{dalvit@df.uba.ar} 
and Francisco D. Mazzitelli
$^{1,2}$\thanks{fmazzi@df.uba.ar}}

\address{$^1$Departamento de F\'{\i}sica ``J.J. Giambiagi'', 
FCEyN, UBA, Pabell\'on 1, Ciudad Universitaria, 1428 Buenos Aires, Argentina}

\address{$^2$Instituto de Astronom\'{\i}a y F\'{\i}sica del Espacio, 
CC67 Suc28, 1428~Buenos~Aires, Argentina}

\maketitle

\begin{abstract}
When graviton loops are
taken into account, the background metric obtained as 
a solution to the one-loop corrected Einstein equations turns out to be 
gauge fixing dependent. Therefore it is of no physical relevance. Instead we 
consider a physical observable, namely the trajectory of a test 
particle in the presence of gravitons.  We derive a quantum
corrected geodesic equation that includes backreaction effects and is 
explicitly independent of any gauge fixing parameter.
\end{abstract}


%%%%%%%%%%%%%%%%%%%%%%%%%%%%%%%%%%%%%%%%%%%%%%%%%%%%%%%%%%%%%%%%%%%%%%%%%%%%
%
%		Main text
%
%%%%%%%%%%%%%%%%%%%%%%%%%%%%%%%%%%%%%%%%%%%%%%%%%%%%%%%%%%%%%%%%%%%%%%%%%%%%

\section{INTRODUCTION}

In quantum field theory there are many physical situations where one is
interested in the dynamical evolution of fields rather 
than in S-matrix elements. The effective action (EA) is a useful tool to
obtain the equations that govern such dynamics including the backreaction
effects due to quantum fluctuations. In the context of gravity,
the equations that give the dynamics of the spacetime metric including
quantum effects are the so-called Semiclassical Einstein Equations
(SEE) \cite{BD}. 
These have been widely used to analyze different physical situations
like gravitational collapse and black hole evaporation.

Since DeWitt's pioneering work \cite{DeWitt}, 
it is known that, at the one-loop
level, the quantization of the fluctuations of the gravitational
field  around a given background is equally as important
as the quantization of the matter fields. Therefore, the 
graviton field contributes to the SEE along with all the other
matter fields.
In order to avoid the technical complications that take place
when gravitons are quantized, their contribution to the SEE 
is usually neglected. It is a common belief that, once the technical
details are solved, one can compute their contribution to the 
energy momentum tensor and write the full one loop SEE. The 
solution to these equations would be the
quantum corrected metric of spacetime.

In the present paper we will argue that, when gravitons are taken into 
account, the solution to the SEE is not physical. 
The reason is simple: any classical device
used to measure the spacetime geometry will also feel the graviton 
fluctuations. As the coupling between the classical device
and the metric is non linear, the device will not measure the `background
geometry' (i.e. the geometry that solves the SEE). As a particular
example we will show that 
a classical particle does not follow a geodesic  of the background
metric. Instead its motion is determined by  a quantum corrected 
geodesic equation that takes into account its coupling to the gravitons.

This analysis also leads us to find a solution to the so-called
gauge fixing problem. A  `technical' obstacle to think of a
solution to the SEE as the metric of spacetime is that in general
it depends on the gauge fixing of the gravitons. As an example
we can mention calculations of compactification radii in Kaluza-Klein
theories \cite{KL}.
The standard approach to tackle this problem is to consider the 
Vilkovisky-DeWitt effective action \cite{V1}, 
which is specifically built to give a reparametrization,
gauge fixing independent action. 
However, this action suffers from another type of
arbitrariness, namely the dependence on the supermetric in the space of fields
that is introduced in its definition \cite{Odintsov,KM,SuperM}.
The aforementioned obstacle is not
`technical' but physical: since the classical device couples to
gravitons, the solution to the SEE will not, in general,
have a clear physical interpretation. We will demonstrate explicitly that while
the solution to the SEE is gauge fixing dependent, the
quantum corrected geodesic equation (that takes into account such
coupling) does not depend on the gauge fixing. 
In summary, the  solution of the backreaction problem consists of two 
steps: to solve the semiclassical Einstein equations and to 
extract the physical quantities from the solution. 

In order to illustrate these facts we will consider the calculation
of the leading quantum corrections to the Newtonian potential. As has been 
pointed out in \cite{D1,D2,WE}, when General Relativity is looked upon as an
effective field theory, low energy quantum effects can be studied without the
knowledge of the (unknown) high energy physics. The leading long distance 
quantum corrections to the gravitational interactions are due to massless 
particles and only involve their coupling at energies low compared to the
Planck mass. Using this idea, many authors have calculated the leading 
quantum corrections to the Newtonian potential computing different sets of
Feynman diagrams \cite{D1,D2,VM,HL}. Instead of evaluating diagrams
and S-matrix elements, we are 
here concerned with a covariant calculation based on EA
and effective field equations. This covariant approach is more
adequate to study problems in which
one considers fluctuations around non-flat backgrounds.
We shall first compute the SEE for the 
backreaction problem starting from the standard EA and show how they depend on 
the gauge fixing. Using a 
corrected geodesic equation we will deduce a physical quantum corrected 
Newtonian potential, which does not depend on the gauge fixing parameters.

%%%%%%%%%%%%%%%%%%%%%%%%%%%%%%%%%%%%%%%%%%%%%%%%%%%%%%%%%%%%%%%%%%%%%%%%%%%%%

\section{THE ONE LOOP EFFECTIVE ACTION FOR GRAVITY+MASS: DIVERGENCES}

The Einstein-Hilbert action for pure gravity is 
\footnote{
Our metric has signature $(-+++)$ and the curvature tensor is defined as
${\bar R}^{\mu}_{\, \cdot \, \nu\alpha\beta} = 
\partial_{\alpha} \Gamma^{\mu}_{\nu\beta} - \ldots$, 
${\bar R}_{\alpha\beta} = {\bar R}^{\mu}_{\, \cdot \, \alpha\mu\beta}$ and
${\bar R}= {\bar g}^{\alpha\beta} {\bar R}_{\alpha\beta}$. We use units
$\hbar=c=1$.
}

\begin{equation}
S_G = \frac{2}{\kappa^2} \int d^4x \sqrt{-{\bar g}} {\bar R} , 
\end{equation}
where ${\bar R}$ is the curvature scalar,
${\bar g}_{\mu\nu}$ is the metric tensor,
${\bar g}={\rm det} {\bar g}_{\mu\nu}$, and $\kappa^2=32 \pi G$, with
$G$ being Newton's constant. In the background field method we consider 
fluctuations of the gravitational field around a background metric,
${\bar g}_{\mu\nu}=g_{\mu\nu} + \kappa s_{\mu\nu}$.
Expanding the action up to quadratic order in the graviton fluctuations 
$s_{\mu\nu}$, the gravitational action reads,

\begin{eqnarray}
S_G &=& \int d^4x \sqrt{-g} \left[ 
\frac{2}{\kappa^2} R + \frac{1}{\kappa} s_{\mu\nu} (g^{\mu\nu}R-2 R^{\mu\nu})
+ 
\left\{ 
\frac{1}{2} \nabla_{\alpha} s_{\mu\nu} \nabla^{\alpha} s^{\mu\nu} -
\frac{1}{2} \nabla_{\alpha}s \nabla^{\alpha} s + 
\nabla_{\alpha} s \nabla_{\beta} s^{\alpha\beta} - \right. \right. \nonumber \\
&&
\left. \left. \nabla_{\alpha} s_{\mu\beta} \nabla^{\beta} s^{\mu\alpha} +
R (\frac{1}{4} s^2 - \frac{1}{2} s_{\mu\nu} s^{\mu\nu}) +
R^{\mu\nu} (2 s^{\lambda}_{\mu} s_{\nu\lambda} - s s_{\mu\nu})
 \right\} + \ldots
\right] ,
\end{eqnarray}
where $s=g^{\mu\nu} s_{\mu\nu}$, and the ellipsis 
denote higher order terms in the fluctuations.
In order to fix the gauge one chooses a gauge fixing function
$\chi^{\mu}[g,s]$, and a gauge fixing action

\begin{equation}
S_{{\rm gf}}[g,s] = -\frac{1}{2} \int d^4x \sqrt{-g}  \chi^{\mu} g_{\mu\nu} 
\chi^{\nu} .
\end{equation}

The one loop effective action for the background metric is obtained from 
integrating out quantum fluctuations and implies the evaluation of 
functional determinants for gravitons and ghosts in the presence of the
background fields. It reads

\begin{equation}
S_{{\rm eff}} = S_G +
\frac{i}{2} {\rm Tr} \ln 
\left[
\frac{\delta^2 S_G[g]}{\delta g^{\alpha\beta} g^{\gamma\delta}} -
\frac{\delta \chi^{\mu}}{\delta g^{\alpha\beta}} g_{\mu\nu}
\frac{\delta \chi^{\nu}}{\delta g^{\gamma\delta}}
\right]
- i {\rm Tr} \ln 
\left[
-2 g_{\sigma\alpha} \nabla_{\beta} 
\frac{\delta \chi^{\mu}}{\delta g^{\alpha\beta}} 
\right] .
\end{equation}
The first term is the classical action, the second one stems from
graviton fluctuations and the last one is the ghosts contribution. These
last two terms are quantum corrections linear in $\hbar$. 

To proceed further one has to choose a particular gauge fixing function. 
The simplest choices of gauge are those called `minimal' gauges, which lead 
to the evaluation of functional traces for gravitons and ghosts of
second-order differential operators of the form 
 $F_{AB}(\nabla) = \hat{C}_{AB} \, g^{\mu\nu} 
 \nabla_{\mu} \nabla_{\nu} + \hat{Q}_{AB}$, where
 $\hat{C}_{AB}$ is an invertible matrix and $\hat{Q}_{AB}$
is an arbitrary matrix.
For these cases the one loop EA can be 
expanded in powers of the background dimensionality using the well-known
Schwinger-DeWitt expansion, which is local in the background fields
(see Appendix A). For the
other `non minimal' gauges, in \cite{BV2} it has been developed a 
reduction method that generalizes the former technique (see Appendix B). 

In the following we shall mainly consider the so-called $\lambda$-family,
which is a one parameter family of gauge fixing functions,

\begin{equation}
\chi^{\mu}(\lambda) = \frac{1}{\sqrt{1+\lambda}} \left[
g^{\mu\gamma} \nabla^{\sigma} s_{\gamma\sigma} - 
\frac{1}{2} g^{\gamma\sigma} \nabla^{\mu} s_{\gamma\sigma} \right] .
\end{equation} 
For gauge fixing functions linear in the metric fluctuations, ghosts decouple
from the fluctuations $s_{\mu\nu}$ and only couple to the background fields.
The one loop EA takes the form
\begin{equation}
S_{{\rm eff}} = S_G +
\frac{i}{2} {\rm Tr} \ln F^{\alpha\beta,\mu\nu}(\nabla) 
- i {\rm Tr} \ln (\Box \delta^{\mu}_{\nu} + R^{\mu}_{\nu}) ,
\end{equation} 
where the second term involves graviton diagrams and the third one involves
ghost diagrams. The second-order differential operator is 

\begin{equation}
F^{\alpha\beta,\mu\nu}(\nabla) = \sqrt{-g} C^{\alpha\beta,\lambda\sigma} 
\left\{ 
\Box \delta^{\mu}_{(\lambda} \delta^{\nu}_{\sigma)} -
\frac{2 \lambda}{1+\lambda} \delta^{(\mu}_{(\lambda} \nabla_{\sigma)}
\nabla^{\nu)} + 
\frac{\lambda}{1+\lambda} g^{\mu\nu} \nabla_{(\lambda} \nabla_{\sigma)}
+ P^{\mu\nu}_{\lambda\sigma}
\right\} ,
\end{equation}
where
\begin{eqnarray}
C^{\alpha\beta,\lambda\sigma} &=& \frac{1}{4} 
(g^{\lambda \alpha} g^{\sigma\beta} + g^{\lambda\beta} g^{\sigma\alpha} -
g^{\lambda \sigma} g^{\alpha \beta} ) , \nonumber \\
P^{\mu\nu}_{\lambda\sigma} &=&
2 R_{\lambda \,\, \cdot \,\, \sigma \cdot}^{~(\mu ~\nu)}
+ 2 \delta^{(\mu}_{(\lambda} R^{\nu)}_{\sigma)}
- g^{\mu\nu} R_{\lambda\sigma} - g_{\lambda\sigma} R^{\mu\nu}
- R \delta^{\mu}_{(\lambda} \delta^{\nu}_{\sigma)}
+ \frac{1}{2} g^{\mu\nu} g_{\lambda\sigma} R . 
\end{eqnarray}
Here the parenthesis denote symmetrization with a $1/2$ factor.
We see that $F$ does not have the form of a minimal 
operator due to the presence of 
the second and third terms. For the special case $\lambda=0$, which is known 
as DeWitt gauge, we have the simplest case of a minimal operator.

Next we couple gravity to a heavy particle (a classical source) of mass $M$, 
which adds a new term to the action
\begin{equation}
S_M= -M \int \sqrt{-{\bar g}_{\mu\nu}dx^{\mu} dx^{\nu} } .
\end{equation}
This coupling introduces 
an additional contribution to the EA. Expanding the action for
the particle up to quadratic order in gravitons, we have
\begin{equation}
S_M = - M \int d\tau \left[
1 - \frac{\kappa}{2} s_{\mu\nu} {\dot x}^{\mu} {\dot x}^{\nu} 
-  \frac{\kappa^2}{8} s_{\mu\nu} s_{\rho\sigma} 
{\dot x}^{\mu}  {\dot x}^{\nu}  
 {\dot x}^{\rho} {\dot x}^{\sigma} + \ldots \right] ,
\end{equation}
where the dots represent derivatives with respect to the proper time $\tau$,
defined as $d\tau^2=-g_{\mu\nu} dx^{\mu} dx^{\nu}$, and the ellipsis are 
higher order terms in the gravitons fluctuations. Introducing an identity as
$1=\int d^4y \sqrt{-g}  \delta^4(y-x(\tau))$, the action can be rewritten
in the following way
\begin{equation}
S_M = -M \int d\tau + 
\frac{\kappa}{2} \int d^4y \sqrt{-g} s_{\mu\nu}(y) T^{\mu\nu}(y) +
\int d^4y \sqrt{-g} s_{\mu\nu}(y) s_{\rho\sigma}(y) 
{\tilde M}^{\mu\nu\rho\sigma}(y) + \ldots ,
\label{eq:smcuad}
\end{equation}
where
\begin{equation}
T^{\mu\nu}(y) = M \int d\tau {\dot x}^{\mu} {\dot x}^{\nu} 
\delta^4(y-x(\tau)) ,
\label{eq:tmunu}
\end{equation}
and
\begin{equation}
{\tilde M}^{\mu\nu\rho\sigma}(y) =\frac{M \kappa^2}{8} 
\int d\tau \delta^4(y-x(\tau))
{\dot x}^{\mu} {\dot x}^{\nu} {\dot x}^{\rho} {\dot x}^{\sigma} .
\end{equation}
The quadratic terms in Eq.(\ref{eq:smcuad}) introduce 
a new contribution to the differential operator
$F(\nabla)$, which finally takes the form
\begin{equation}
F^{\alpha\beta,\mu\nu}(\nabla) = 
\sqrt{-g} C^{\alpha\beta,\lambda\sigma} 
\left\{ 
\Box \delta^{\mu}_{(\lambda} \delta^{\nu}_{\sigma)} -
\frac{2 \lambda}{1+\lambda} \delta^{(\mu}_{(\lambda} \nabla_{\sigma)}
\nabla^{\nu)} + 
\frac{\lambda}{1+\lambda} g^{\mu\nu} \nabla_{(\lambda} \nabla_{\sigma)}
+ P^{\mu\nu}_{\lambda\sigma} + M^{\mu\nu}_{\lambda\sigma}
\right\} ,
\end{equation}
with
\begin{equation}
M^{\mu\nu}_{\lambda\sigma}(y) = (C^{-1})^{\mu\nu\alpha\beta} 
{\tilde M}_{\alpha\beta\lambda\sigma}(y)
= \frac{M \kappa^2}{8} \int d\tau \delta^4(y-x(\tau)) 
\left[ g^{\mu\nu} {\dot x}_{\lambda} {\dot x}_{\sigma} +
2 {\dot x}^{\mu} {\dot x}^{\nu} {\dot x}_{\lambda} {\dot x}_{\sigma} 
\right] .
\label{eq:meff}
\end{equation}

As is well-known, the EA has divergences. For example, for the
pure gravitational part, the one loop divergences in the DeWitt
($\lambda =0$) gauge have been calculated long ago using dimensional 
regularization and turn out to be local terms quadratic in the curvature
tensors \cite{THV}. They read
\footnote{
To be precise, the EA contains two additional divergences, one proportional
to $\sqrt{-g} $ and another proportional to $\sqrt{-g} R$. As these can be
absorbed into a redefinition of the cosmological constant and the Newton
constant, we shall not consider them in what follows.}

\begin{eqnarray}
\Delta S_G^{\rm div}(\lambda=0) &=& 
\frac{2}{(4-d) 96 \pi^2} \int d^4x \sqrt{-g}
\left[ 
\frac{53}{15} (R_{\mu\nu\rho\sigma} R^{\mu\nu\rho\sigma} - 
4 R_{\mu\nu} R^{\mu\nu} + R^2) + \right. \nonumber \\
&& \left. ~~~~~~~~~~~~~~~~~~~~~~~~~~~~~~~~~~~~
\frac{21}{10} R_{\mu\nu} R^{\mu\nu} +
\frac{1}{20} R^2 
\right] ,
\label{eq:divergenciesG}
\end{eqnarray}
where the first term in brackets is the Gauss-Bonnet term, a topological
invariant in $d=4$ spacetime dimensions. In Appendix A we show how to
evaluate the divergence stemming from the massive part for the
minimal gauge $\lambda=0$. It reads

\begin{eqnarray}
\Delta S_M^{\rm div}(\lambda=0) &=&
 \frac{2}{(4-d) 64 \pi^2} \int d^4x \sqrt{-g}
\left[
M_{\mu\nu\rho\sigma} M^{\mu\nu\rho\sigma} + \right. \nonumber \\
&& \left. ~~~~~~~~~~~~~~~~~~~~~~~~~~~~~~~~
2  M_{\mu\nu\rho\sigma}  
\left( P^{\rho\sigma\mu\nu} + \frac{1}{6} R \delta^{\rho(\mu} 
\delta^{\sigma\nu)} \right) \right] .
\label{eq:divergenciesM}
\end{eqnarray} 

Now we have to calculate the EA for any member of the $\lambda$-family
gauge fixing funtions other than the $\lambda=0$ one. The calculation
is cumbersome and we leave it for Appendix B.
Here we just state the main result that shall concern us (see below), 
namely the divergence of the one loop EA that is linear in the extremal
${\cal E}^{\mu\nu}= 
-\frac{2}{\kappa^2} (R^{\mu\nu}-\frac{1}{2} R g^{\mu\nu})
+\frac{1}{2} T^{\mu\nu}$,

\begin{equation}
\Delta S^{\rm div}(\lambda) = \Delta S^{\rm div}(\lambda=0)
- \frac{\lambda}{4-d} \, \frac{\kappa^2}{24 \pi^2} \int d^4x \sqrt{-g}
\left[ -5 R_{\mu\nu} {\cal E}^{\mu\nu} +
\frac{5}{2} R g_{\mu\nu}  {\cal E}^{\mu\nu} \right] ,
\label{eq:divergenciesL}
\end{equation}
where 
 $\Delta S^{\rm div}(\lambda=0)= \Delta S^{\rm div}_G(\lambda=0) + 
 \Delta S^{\rm div}_M(\lambda=0)$ 
is the divergence for the DeWitt gauge,
that was already calculated. Note that the (ultraviolet) divergences of the 
EA take the form of local tensors expressed in terms of curvatures and the
energy-momentum tensor for the source particle.

%%%%%%%%%%%%%%%%%%%%%%%%%%%%%%%%%%%%%%%%%%%%%%%%%%%%%%%%%%%%%%%%%%%%%%%%

\section{LONG DISTANCE LEADING QUANTUM CORRECTIONS: THE LOG TERMS}

The theory we are considering is not renormalizable, since the 
divergences cannot be absorbed into the parameters introduced thus far.
Additional divergent counterterms (and some accompanying finite parts) 
quadratic in the curvature tensors must be added to the classical action
$S_G+S_M$.
However, the nonrenormalizability of the theory is not an impediment for
making well defined quantum predictions at low energies/large distancies. As
we have already remarked, the idea is to treat gravity as an effective field
theory, and perform a systematic expansion in the energy.
In this approach, the unknown parameters 
introduced with the various counterterms have to be determined by comparison
with experiment, which then allows to make predictions to a given order in
an energy expansion. However, the low energy physics is not contained in these
parameters, but rather in a different class of quantum corrections. 
The leading long distance corrections stem from the non-local, non-analytic 
terms in the one loop effective action. These non-local terms have been
computed in \cite{Gospel,Avra}
expanding the EA in powers of the curvatures, using
a resummation procedure of the Schwinger-DeWitt expansion for the action.
Keeping up to quadratic order in the curvature tensors, the general form of 
such terms is ${\cal R} G(\Box) {\cal R}$, where ${\cal R}$ denotes any 
of the tensors $R, R_{\mu\nu},M_{\mu\nu\rho\sigma}$, and 
$G(\Box)$ is a non-local form factor. For the theory we
are considering, $G(\Box)$ is proportional to $\ln(-\Box)$, and these
logarithmic terms are the relevant ones in the low energy limit. The
proportionality constants accompanying the $\ln(-\Box)$ can be read off from 
the (local) divergences in 
Eqs.(\ref{eq:divergenciesG},\ref{eq:divergenciesM},\ref{eq:divergenciesL}) 
in a manner outlined in \cite{D2,Gospel}. One extracts the coefficient
of the logarithmic correction from the divergence in the following way:

\begin{equation}
\frac{\alpha}{4-d} \int d^4x \sqrt{-g} (\ldots) \rightarrow
- \frac{\alpha}{2} \int d^4x \sqrt{-g} (\ldots) \ln(-\Box) .
\end{equation}
Using this result, the non-local part of the EA proportional to the 
logarithm takes the form 
 $\Delta S = \Delta S^{{\rm nl}}_G(\lambda = 0) 
 + \Delta S^{{\rm nl}}_M(\lambda = 0)  + 
 \Delta S^{{\rm nl}}(\lambda \neq 0)$, with

\begin{eqnarray}
\Delta S^{{\rm nl}}_G(\lambda=0) &=&
- \frac{1}{96 \pi^2} \int d^4x \sqrt{-g}  \left[  \frac{21}{10} R_{\mu\nu} 
\ln(-\Box) R^{\mu\nu} +  \frac{1}{20} R \ln(-\Box) R \right] , \\
\Delta S_M^{{\rm nl}}(\lambda=0) &=& -\frac{1}{64 \pi^2} \int d^4x \sqrt{-g}
\left[ M_{\mu\nu\rho\sigma} \ln(-\Box)  M^{\rho\sigma\mu\nu} + \right.
\nonumber \\
&& \left. ~~~~~~~~~~~~~~~~~~~~~~~~
2  M_{\mu\nu\rho\sigma} \ln(-\Box) \left( P^{\rho\sigma\mu\nu} +
\frac{1}{6} R \delta^{\rho(\mu} \delta^{\sigma\nu)} \right) \right] , 
\label{eq:nolocM}
\end{eqnarray}
\begin{eqnarray}
\Delta S^{{\rm nl}}(\lambda \neq 0) &=&
\int d^4x \sqrt{-g} 
\left[ a(\lambda) R_{\mu\nu} \ln(-\Box) {\cal E}^{\mu\nu} +
b(\lambda) R g_{\mu\nu} \ln(-\Box) {\cal E}^{\mu\nu} \right] ,
\label{eq:nolocL}
\end{eqnarray} 
where $a(\lambda)=-\frac{5 \lambda \kappa^2}{48 \pi^2}$ and
$b(\lambda)=\frac{5 \lambda \kappa^2}{96 \pi^2}$.
 
We choose a classical static point mass located at the origin. Hence
 ${\dot x}^{\mu} =(1,0,0,0)$, 
 $T^{\mu\nu}(x)=M \delta^{\mu}_0 \delta^{\nu}_0 \delta^3({\vec x})$
and $T^{\mu}_{\mu}=-M \delta^3({\vec x})$.
As we will calculate long distance corrections 
to gravitational interactions (in particular to the Newtonian potential), 
we can assume the source is a `point mass', although its size should be 
much larger than its Schwarzschild radius and the Planck length in order to
justify the weak field approximation to be done in what follows. With this 
choice for the source, the different tensors appearing in the massive non-local
part of the EA take the form

\begin{eqnarray}
M^{\mu\nu\lambda\sigma}(y) &=&
\frac{M \kappa^2}{8} \delta^3({\vec y}) \,
[g^{\mu\nu}+2 \delta^{\mu}_0 \delta^{\nu}_0] \, \delta^{\lambda}_0 
\delta^{\sigma}_0 , \nonumber\\
M_{\mu\nu\rho\sigma}  R \delta^{\rho(\mu} \delta^{\sigma\nu)} &=&
\frac{M \kappa^2}{8} R  \delta^3({\vec y}), \\
M_{\mu\nu\rho\sigma} P^{\rho\sigma\mu\nu} &=&
\frac{M \kappa^2}{8}  \delta^3({\vec y}) 
[g^{\mu\nu} P_{00\mu\nu} + 2 P_{0000}]=
-  \frac{M \kappa^2}{8} R  \delta^3({\vec y}). \nonumber
\end{eqnarray}
With the help of these expressions, the contribution of the source to the
nonlocal part of the EA is

\begin{equation}
\Delta S_M^{{\rm nl}} (\lambda=0) = \frac{5 M \kappa^2}{1536 \pi^2} \int d^4x 
\sqrt{-g} R \ln(-\nabla^2) \delta^3({\vec x}) ,
\end{equation}
where we have used the fact that the mass $M$ is static to replace
$\Box \rightarrow \nabla^2$. We have omitted the term that is quadratic
in $M$ because it will be irrelevant in the long distance limit.

Adding the classical and quantum contributions of the EA and taking functional
derivations with respect to the metric, it is possible to compute the 
SEE including backreaction of gravitons.
As we are neglecting ${\cal O}({\cal R}^3)$ terms in the effective action, 
it makes no sense to retain 
${\cal O}({\cal R}^2)$ terms in the equations of motion. Therefore, when
doing the variation of the action with respect to the metric, it is not
necessary to take into account the $g_{\mu\nu}$ dependence of the logarithmic
form factors. Moreover it is possible to commute the covariant derivatives
acting on a curvature, i.e., 
$\nabla_{\mu} \nabla_{\nu} {\cal R} = 
\nabla_{\nu} \nabla_{\mu} {\cal R} + {\cal O} ({\cal R}^2)$.
However, if one uses the standard in-out EA calculated thus far, the
equations of motion turn out to be neither real nor causal. In order to
get the equations for the mean values one can take any of
the following routes: to calculate the in-in EA (which involves a
doubling of the number of fields) and derive from it the appropiate
field equations \cite{CTP}, to take twice the real and causal part of the
in-out equations, or to calculate the euclidean EA and replace in the
equations of motion the euclidean propagators by the retarded ones \cite{DM}.
Using any of these alternatives, the mean value equations, up to linear order
in curvatures, read,

\begin{equation}
\frac{1}{8\pi G} (R_{\mu\nu}-\frac{1}{2} R g_{\mu\nu}) =
T_{\mu\nu} + \langle T_{\mu\nu}\rangle^{G}_{\lambda=0} +
\langle T_{\mu\nu}\rangle^{M}_{\lambda=0} +
\langle T_{\mu\nu}\rangle_{a(\lambda)} +
\langle T_{\mu\nu}\rangle_{b(\lambda)} ,
\end{equation}
where

\begin{eqnarray}
\langle T_{\mu\nu}\rangle^{G}_{\lambda=0} &=&
-\frac{1}{96 \pi^2} \left[ \frac{21}{10} \ln(-\nabla^2) H_{\mu\nu}^{(2)} +
\frac{1}{20} \ln(-\nabla^2) H_{\mu\nu}^{(1)} \right] ,\nonumber\\
\langle T_{\mu\nu}\rangle^{M}_{\lambda=0} &=&
\frac{5 M \kappa^2}{768 \pi^2} (\nabla_{\mu} \nabla_{\nu} - g_{\mu\nu} 
\nabla^2) \ln(-\nabla^2) \delta^3({\vec x}) ,\nonumber\\
\langle T_{\mu\nu}\rangle_{a(\lambda)} &=&
a(\lambda) \left[-\frac{2}{\kappa^2} \ln(-\nabla^2) H_{\mu\nu}^{(2)} +
\frac{1}{\kappa^2}  \ln(-\nabla^2) H_{\mu\nu}^{(1)} -\frac{1}{2}
\nabla^2 \ln(-\nabla^2) T_{\mu\nu} \right] ,\nonumber\\
\langle T_{\mu\nu}\rangle_{b(\lambda)} &=&
b(\lambda) \left[ \frac{2}{\kappa^2} \ln(-\nabla^2) H_{\mu\nu}^{(1)}
+ \nabla_{\mu} \nabla_{\nu} \ln(-\nabla^2) T^{\alpha}_{\alpha} -
g_{\mu\nu} \nabla^2 \ln(-\nabla^2) T^{\alpha}_{\alpha} \right] ,
\label{eq:tensors}
\end{eqnarray}
where we have introduced the tensors 
$H_{\mu\nu}^{(1)}=4\nabla_{\mu} \nabla_{\nu} R-4g_{\mu\nu} \nabla^2 R$ 
and 
$H_{\mu\nu}^{(2)}=2\nabla_{\mu} \nabla_{\nu} R-g_{\mu\nu} \nabla^2 R -
2 \nabla^2 R_{\mu\nu}$. The non-local operator $\ln(-\nabla^2)$ acts on the 
delta function as 
\footnote
{
This expression can also be obtained by means of the Fourier transform
 $\int \frac{d^3q}{(2\pi)^3} e^{- i {\vec q} \cdot {\vec r}} \ln q^2
 = - \frac{1}{2 \pi r^3}$.
}
$\ln(-\nabla^2) \delta^3({\vec x}) = -\frac{1}{2\pi r^3}$ 
\cite{DM}. 

%%%%%%%%%%%%%%%%%%%%%%%%%%%%%%%%%%%%%%%%%%%%%%%%%%%%%%%%%%%%%%%%%%%%%%%%%%%

\section{QUANTUM CORRECTIONS TO THE CLASSICAL METRIC}

In order to solve the effective Einstein equations for the background metric
we shall make perturbations around flat spacetime,
$g_{\mu\nu}=\eta_{\mu\nu}+h_{\mu\nu}$ with $\eta_{\mu\nu}={\rm diag}(-+++)$. 
We choose the harmonic gauge 
 $(h_{\mu\nu} - \frac{1}{2} h \eta_{\mu\nu})^{;\nu}=0$
for the background perturbation metric. 
It is worth  mentioning that this choice is completely independent of 
the gauge fixing problem for the quantum fluctuations. In this gauge, the Ricci
tensor is $R_{\mu\nu}=-\frac{1}{2} \nabla^2 h_{\mu\nu}$ and the Ricci scalar
$R=-\frac{1}{2} \nabla^2 h$, with $h=\eta^{\mu\nu} h_{\mu\nu}$. Indeces
are lowered and raised with the flat metric. The equations of motion take the
form
\begin{equation}
\nabla^2 {\bar h}_{\mu\nu} = -16 \pi G \left[
T_{\mu\nu} + \langle T_{\mu\nu}\rangle^{G}_{\lambda=0} +
\langle T_{\mu\nu}\rangle^{M}_{\lambda=0} +
\langle T_{\mu\nu}\rangle_{a(\lambda)} +
\langle T_{\mu\nu}\rangle_{b(\lambda)} \right] ,
\label{eq:htecho}
\end{equation}
where ${\bar h}_{\mu\nu} = h_{\mu\nu} - \frac{1}{2} h \eta_{\mu\nu}$. The terms
in the rhs are those appearing in Eq.(\ref{eq:tensors}) 
evaluated in the weak field 
approximation. In this approximation, the Newtonian potential is related to
the $00$-component of the perturbation metric as $V(r)=-\frac{1}{2} h_{00}$.
In order to find $h_{00}$ we solve Eq.(\ref{eq:htecho}) 
for ${\bar h}_{00}$ and the trace
of that equation for $h$. We find a perturbative solution to these equations
around the classical solutions. This perturbative approach is the reason for 
having omitted terms in the EA that are proportional to the square of the 
extremal ${\cal E}^{\mu\nu}$. These contribute to the rhs of 
Eq.(\ref{eq:htecho})
with terms proportional to the classical equations, and therefore vanish
identically when the equations are solved perturbatively. We write 
${\bar h}_{00} = {\bar h}_{00}^{(0)} + {\bar h}_{00}^{(1)}$, where
${\bar h}_{00}^{(0)} = \frac{4 G M}{r}$ is the classical contribution, and
${\bar h}_{00}^{(1)}$ is the quantum correction. We get

\begin{equation}
{\bar h}_{00}^{(1)} = - \frac{2}{15 \pi} \frac{G^2 M}{r^3} +
\frac{5}{3 \pi} \frac{G^2 M}{r^3} + 4 a(\lambda) \frac{G M}{r^3} +
8 b(\lambda) \frac{G M}{r^3} ,
\end{equation}
where the first and second terms come from the pure gravitational and massive
part of the EA (for the DeWitt $\lambda=0$ gauge) and the last two terms 
correspond to other gauges of the $\lambda$-family. 
The equation for the trace is

\begin{equation}
\nabla^2 h = 16 \pi G  
\left[
T^{\mu}_{\mu} + \langle T^{\mu}_{\mu}\rangle^{G}_{\lambda=0} +
\langle T^{\mu}_{\mu}\rangle^{M}_{\lambda=0} +
\langle T^{\mu}_{\mu}\rangle_{a(\lambda)} +
\langle T^{\mu}_{\mu}\rangle_{b(\lambda)}
\right] ,
\end{equation}
whose perturbative solution $h=h^{(0)}+h^{(1)}$ leads to a classical 
term $h^{(0)}=\frac{4 G M}{r}$ and a quantum correction

\begin{equation}
h^{(1)} = -\frac{18}{3 \pi} \frac{G^2 M}{r^3} + 
\frac{5}{\pi} \frac{G^2 M}{r^3} + 4 a(\lambda) \frac{G M}{r^3} 
+ 24 b(\lambda)  \frac{G M}{r^3} .
\end{equation}
The origin of each term is the same as previously discussed. Therefore
the $00$-component of the perturbation $h_{\mu\nu}$ reads

\begin{equation}
h_{00} = {\bar h}_{00} - \frac{1}{2} h  =
 \frac{2 G M}{r} \left[ 1 + \frac{43 G}{30 \pi r^2} - \frac{5 G}{12 \pi r^2} +
\frac{ a(\lambda) - 2 b(\lambda)}{r^2} \right] . 
\label{eq:BRmetric}
\end{equation}
The first term is due to the presence of the classical mass M (for simplicity
we consider only the Newtonian limit, that is, we do not include classical
corrections from general relativity). The last four terms are quantum
corrections. The second one stems from pure gravitational 
contributions (vacuum polarization)
while the remaining ones arise from the coupling of the mass $M$ to gravitons.
The Newtonian potential follows through the identity
 $V(r)=- \frac{1}{2} h_{00}$. 
We stress again that the non-local logarithmic corrections to the effective
action give the leading quantum corrections in the long distance limit,
that are proportional to $r^{-3}$. Had we considered additional terms
proportional to ${\cal R}^2$ in the effective action, we would had obtained
additional corrections to the classical metric that vanish exponentially as
$r \rightarrow \infty$.

From Eq.(\ref{eq:BRmetric}) it is then clear that the metric 
that solves the backreaction equations for the one loop quantized gravity 
depends on which particular function one chooses to fix the gauge. It is for 
this reason that the classical geodesic equation for such metric 
cannot be physical. 

%%%%%%%%%%%%%%%%%%%%%%%%%%%%%%%%%%%%%%%%%%%%%%%%%%%%%%%%%%%%%%%%%%%%%%%%

\section{QUANTUM CORRECTED GEODESIC EQUATION}

Let us consider a classical test particle 
of mass $m$ in
the presence of the quantized gravitational field 
${\bar g}_{\mu\nu}$. A physical observable should be the motion of 
this particle. 
We consider that the mass of this particle is much smaller than $M$, which 
allows us to neglect all contributions of the test particle to the solution
Eq.(\ref{eq:BRmetric}) of the one loop corrected equation.
Now comes the key ingredient: in order to determine how this test 
particle moves, one also has to take into account the fact that it couples to 
the quantum metric ${\bar g}_{\mu\nu}$ through the term 
 $-m \int \sqrt{- {\bar g}_{\mu\nu}(z) dz^{\mu} dz^{\nu} }$, where $z^{\mu}$ 
denotes the path of the test particle. Therefore there will be an extra 
contribution to the one loop EA due to this coupling to gravitons, which
in turn will introduce a correction to the geodesic equation. 
It can be obtained from Eqs.(\ref{eq:nolocM},\ref{eq:nolocL}) replacing
 $M_{\mu\nu\rho\sigma}$ by $m_{\mu\nu\rho\sigma} + M_{\mu\nu\rho\sigma}$
and $T^{\mu\nu}$ by  $T^{\mu\nu} + T^{\mu\nu}_m$ and keeping terms linear
in $m$.   
Here the tensor $m_{\mu\nu\rho\sigma}$ is 
the one given in Eq.(\ref{eq:meff}) with $M$ replaced by $m$ and
 $x_{\mu}$ replaced by $z_{\mu}$, and
$T^{\mu\nu}_m$ is the energy-momentum tensor
for the test particle, given in Eq.(\ref{eq:tmunu}), with the same 
replacement. This contribution is

\begin{eqnarray}
\Delta S_m &=& \int d^4x \sqrt{-g}
\left[ 
-\frac{1}{32 \pi^2} m_{\mu\nu\rho\sigma} \ln(-\Box) M^{\rho\sigma\mu\nu} - 
\right. \nonumber \\
&& \left. ~~~~~~~~~~~~~~~~~
\frac{1}{32 \pi^2}  m_{\mu\nu\rho\sigma} \ln(-\Box) 
\left( P^{\rho\sigma\mu\nu} +
\frac{1}{6} R \delta^{\rho(\mu} \delta^{\sigma\nu)} \right) + \right.
\nonumber\\
&& \left. ~~~~~~~~~~~~~~~~~ 
\frac{a(\lambda)}{2} R_{\mu\nu} \ln(-\Box) T^{\mu\nu}_m + 
   \frac{b(\lambda)}{2} R g_{\mu\nu} \ln(-\Box) T^{\mu\nu}_m \right] ,
\label{eq:eseeme}
\end{eqnarray}
The first two terms correspond to the $\lambda=0$ gauge fixing, and the last
two are extra terms appearing for any other gauge. 

The geodesic equation for the test particle can be obtained by taking the 
functional derivative of the effective action with respect to the coordinates
of the particle

\begin{equation}
0=\frac{1}{m} \frac{\delta S_{{\rm eff}}}{\delta z_{\rho}}
= - \left[
\frac{d^2 z^{\rho}}{d \tau^2} + \Gamma^{\rho}_{\mu\sigma}
\frac{d z^{\mu}}{d \tau} \frac{d z^{\sigma}}{d \tau} \right]
+\frac{1}{m} \frac{\delta \Delta S_m}{\delta z_{\rho}} ,
\end{equation}
where $\Gamma^{\rho}_{\mu\sigma}$ is the Christoffel symbol and 
$d\tau^2=-g_{\mu\nu} dz^{\mu} dz^{\nu}$. In the weak, nonrelativistic 
Newtonian limit, the quantum corrected geodesic equation reads

\begin{equation}
\frac{d^2{\vec z}}{dt^2} - \frac{1}{2} {\vec \nabla}h_{00} =
\frac{1}{m} \frac{\delta \Delta S_m}{\delta {\vec z}} .
\end{equation}
Note that $h_{00}$, given in Eq. (\ref{eq:BRmetric}), depends on $a(\lambda)$ 
and $b(\lambda)$.

Now we proceed to evaluate the rhs of this equation. To that end we first 
calculate the different terms in $\Delta S_m$. Using the expression for
$M^{\mu\nu\rho\sigma}$ corresponding to the static source, the first term
of Eq.(\ref{eq:eseeme}) reads

\begin{eqnarray}
&&\Delta S_{m,M}(\lambda=0) \equiv 
-\frac{1}{32 \pi^2} \int d^4y \sqrt{-g} m_{\mu\nu\rho\sigma} \ln(-\Box)
M^{\rho\sigma\mu\nu} = \nonumber \\
&& -\frac{1}{32 \pi^2} \frac{m M \kappa^2}{64} \int d^4y \sqrt{-g} 
\ln(-\Box) \delta^3({\vec y}) \int d\tau \delta^4(y-z(\tau)) 
[2 {\dot z}_0 {\dot z}_0 + 2 g_{00} {\dot z}_0 {\dot z}_0 +
4 {\dot z}_0 {\dot z}_0 {\dot z}_0 {\dot z}_0 ] \approx \nonumber\\
&& - \frac{m M  \kappa^4}{512 \pi^2} \int d\tau \ln(-\Box) 
\delta^3(z(\tau)) .
\label{eq:mM}
\end{eqnarray}
Here we have used the fact that, in the non relativistic limit,
 ${\dot z}_0 \approx -1$. As $\Delta S_m$ is proportional to $\hbar$,
we have also set the metric
$g_{\mu\nu}$ in this equation equal 
to the classical one $\eta_{\mu\nu}$ - any other correction would contribute
with terms ${\cal O}(\hbar^2)$. 
In a similar fashion

\begin{eqnarray}
\Delta S_{m,R}(\lambda=0) & \equiv & 
- \frac{1}{192 \pi^2} \int d^4y \sqrt{-g} m_{\mu\nu\rho\sigma} \ln(-\Box)
R \delta^{\rho(\mu} \delta^{\sigma\nu)} = \nonumber\\
&& - \frac{m \kappa^2}{1536 \pi^2} \int d\tau \ln(-\Box) R(z(\tau)) .
\end{eqnarray}
The other terms appearing in Eq.(\ref{eq:eseeme}) are 

\begin{eqnarray} 
\Delta S_{m,P}(\lambda=0) & \equiv & 
- \frac{1}{32 \pi^2} \int d^4y \sqrt{-g} m^{\mu\nu\rho\sigma} \ln(-\Box) 
P_{\rho\sigma\mu\nu} = \nonumber\\
&& - \frac{m \kappa^2}{256 \pi^2} \int d\tau 
[g^{\mu\nu} {\dot z}^{\rho}  {\dot z}^{\sigma} + 2
{\dot z}^{\mu} {\dot z}^{\nu} {\dot z}^{\rho} {\dot z}^{\sigma}]
\ln(-\Box) P_{\rho\sigma\mu\nu} ,
\end{eqnarray}

\begin{equation}
\Delta S_{m,a(\lambda)}  \equiv 
\frac{a(\lambda)}{2} \int d^4y \sqrt{-g} 
R_{\mu\nu} \ln(-\Box) T^{\mu\nu}_m = 
a(\lambda) \frac{m}{2} \int d\tau {\dot z}^{\mu}{\dot z}^{\nu} 
\ln(-\Box) R_{\mu\nu}(z(\tau)) , 
\end{equation} 
and

\begin{equation}
\Delta S_{m,b(\lambda)} \equiv 
\frac{b(\lambda)}{2} \int d^4y \sqrt{-g} 
R g_{\mu\nu} \ln(-\Box) T^{\mu\nu}_m = 
- b(\lambda) \frac{m}{2} \int d\tau \ln(-\Box) R(z(\tau)) .
\end{equation}
In these equations the Ricci scalar in the one for the classical metric, i.e.
 $R(z(\tau))=-\frac{1}{2} \nabla^2 h^{(0)}(z(\tau))=8\pi G M 
  \delta^3({\vec z}(\tau))$. The same holds for the Ricci tensor 
 $R_{\mu\nu}(z(\tau))$. 

Now we take the variation with respect to ${\vec z}$. We obtain

\begin{eqnarray}
\frac{\delta}{\delta {\vec z}} \Delta S_{m,M}(\lambda=0)  &=& 
\frac{m M G^2}{\pi} {\vec \nabla}(\frac{1}{r^3}) , \nonumber\\
\frac{\delta}{\delta {\vec z}} \Delta S_{m,R}(\lambda=0)  &=& 
 \frac{m M G^2}{12 \pi} {\vec \nabla}(\frac{1}{r^3}) , \nonumber\\
\frac{\delta}{\delta {\vec z}} \Delta S_{m,P}(\lambda=0)  &=& 
- \frac{m M G^2}{2 \pi} {\vec \nabla}(\frac{1}{r^3}) , \nonumber\\
\frac{\delta}{\delta {\vec z}} \Delta S_{m,a(\lambda)} &=& 
- a(\lambda) m M G  {\vec \nabla}(\frac{1}{r^3}) , \nonumber\\
\frac{\delta}{\delta {\vec z}} \Delta S_{m,b(\lambda)} &=& 
2 b(\lambda) m M G  {\vec \nabla}(\frac{1}{r^3}) ,
\end{eqnarray}
where $r=|{\vec z}|$. Therefore

\begin{equation}
\frac{d^2 {\vec z}}{dt^2} - \frac{1}{2} \vec{\nabla} h_{00} = 
\frac{1}{m} \frac{\delta \Delta S_m}{\delta {\vec z}} = 
\left[ \frac{7 G}{12 \pi} - a(\lambda) + 2 b(\lambda) \right]  
{\vec \nabla} \left( \frac{G M}{r^3} \right) .
\end{equation} 
Inserting Eq.(\ref{eq:BRmetric}) into this expression we see that
those gauge fixing dependent terms arising from the backreaction metric 
cancel exactly those coming from the coupling of the test particle to
gravitons. Note that the terms with $a(\lambda)$ and $b(\lambda)$ cancel
separately. 

One can perform the same calculation as before for any gauge not belonging
to the $\lambda$-family in a straightforward manner. As it was already 
mentioned, the difference between the EA for the $\lambda=0$ gauge and that
for any other gauge must be proportional to the extremal ${\cal E}^{\mu\nu}$,
which vanishes on shell. Keeping up to quadratic order in curvature, this
requirement fixes the most general form such a difference can have
(we concentrate on the non-analytic log terms)

\begin{equation}
\Delta S|_{\rm given \, gauge} - \Delta S(\lambda=0) =
\int d^4x \sqrt{-g} [a R_{\mu\nu} \ln(-\Box) {\cal E}^{\mu\nu} +
b R g_{\mu\nu} \ln(-\Box) {\cal E}^{\mu\nu} +
{\cal O} ({\cal E}^{\mu\nu})^2 ] ,
\end{equation}
where $a$ and $b$ are constants that depend on which particular gauge one uses.
For example, for the $\lambda$-family, 
$a=a(\lambda)=- 5 \lambda \kappa^2/48 \pi^2$ and  
$b=b(\lambda)= 5 \lambda \kappa^2/96 \pi^2$, as we have already seen.
In view of the above calculations, we conclude that
the cancelation of the $a$ and $b$ dependent terms takes place for any
possible gauge fixing. 
In this way we obtain a physical, gauge fixing independent 
Newtonian potential $V(r)$ which we read from 
 $d^2 {\vec z}/dt^2 = - {\vec \nabla} V$, namely
   
\begin{equation}
V(r) = -\frac{G M}{r} \left[1 + \frac{43 G\hbar}{30 \pi r^2 c^3} - 
\frac{5 G \hbar}{12 \pi r^2 c^3} +\frac{7 G \hbar}{12 \pi r^2 c^3} \right] ,
\label{eq:potnew2}
\end{equation}
where we have restored units ($\hbar$ and $c$).
A comparison between Eq.(\ref{eq:BRmetric}) and Eq.(\ref{eq:potnew2})
shows that the coupling of the test particle with the gravitons produces
an additional contribution to the Newtonian potential (the last term
in Eq.(\ref{eq:potnew2})) and makes it gauge fixing independent.
Note that the long distance
quantum correction above is extremely small to be measured. However the
specific number is less important than the conceptual fact that the potential
and motion of the test particle are gauge fixing independent.  

%%%%%%%%%%%%%%%%%%%%%%%%%%%%%%%%%%%%%%%%%%%%%%%%%%%%%%%%%%%%%%%%%%%%%%%%%

\section{CONCLUSIONS}

We hope to have convinced the reader that if she/he is interested in
solving the backreaction problem including the graviton contribution,
it is not enough to solve the semiclassical Einstein equations because
they are gauge fixing dependent and not physical. Rather she/he has to
look for physical observables. As an illustration of this point we have
chosen the trajectory of a test particle and we have explicitly shown that,
in the Newtonian limit, the usual effective action gives a gauge fixing 
independent result. 

We would like to mention several lines for future research. On the one hand,
it is of interest to check whether the Newtonian effective potential 
derived in this paper does not depend on reparametrizations of the variables
chosen to perform the perturbative expansion. When working within the 
Vilkovisky-DeWitt approach, the potential should not depend on the 
supermetric defined on the space of fields. On the other hand, it would
be interesting to find the quantum corrected geodesic equation in a 
cosmological setting (desirably, beyond the Newtonian approximation).

Finally, we would like to point out that similar ideas to the one proposed 
here can be applied to the analysis of the mean value equations of any gauge 
theory, for example when computing gluon backreaction effects on classical 
solutions to Yang Mills theories.

%%%%%%%%%%%%%%%%%%%%%%%%%%%%%%%%%%%%%%%%%%%%%%%%%%%%%%%%%%%%%%%%%%%%%%%%%%%%
%
%            Acknowledgements
%
%%%%%%%%%%%%%%%%%%%%%%%%%%%%%%%%%%%%%%%%%%%%%%%%%%%%%%%%%%%%%%%%%%%%%%%%%%%%

\acknowledgements

We acknowledge the support  
from UBACyT, Fundaci\'on Antorchas and Conicet (Argentina).
F.D.M. thanks Jorge Russo for useful discussions on related matters.

%%%%%%%%%%%%%%%%%%%%%%%%%%%%%%%%%%%%%%%%%%%%%%%%%%%%%%%%%%%%%%%%%%%%%%%%%

\appendix
\section{DIVERGENCES FOR MINIMAL GAUGES}

In this Appendix we calculate the divergence of the one-loop EA for
the DeWitt gauge $\lambda=0$. We follow closely the methods thoroughly
explained in \cite{BV2}. 
For DeWitt's gauge, the second-order
differential operator $F(\nabla)$ is 
\begin{equation}
F^{\alpha\beta,\mu\nu}(\nabla|\lambda=0) = 
\sqrt{-g} C^{\alpha\beta,\lambda\sigma} 
\left\{ 
\Box \delta^{\mu}_{(\lambda} \delta^{\nu}_{\sigma)} 
+ P^{\mu\nu}_{\lambda\sigma} + M^{\mu\nu}_{\lambda\sigma}
\right\} ,
\end{equation}
and the one loop EA has the following expression,
\begin{equation}
S_{{\rm eff}} = S_{\rm class}+
\frac{i}{2} {\rm Tr} \ln F^{\alpha\beta,\mu\nu}(\nabla) 
- i {\rm Tr} \ln (\Box \delta^{\mu}_{\nu} + R^{\mu}_{\nu}) ,
\end{equation} 
the first term being the classical action. In the gauge under consideration,
both the differential operator
for the gravitons and the one for the ghosts have a minimal form,
which in matrix notation reads
\begin{equation}
\hat{\cal F}(\nabla) = \Box + \hat{\cal Q} - \frac{1}{6} R \hat{1} .
\end{equation}
Indeed, for the gravitons the matrix $\hat{\cal Q}$ is given by 
 $\hat{\cal Q}= \hat{P} + \hat{M} + \frac{1}{6} R \hat{1}$, while
for the ghosts $\hat{\cal Q}= \hat{R} + \frac{1}{6} R \hat{1}$. 
In order to calculate the functional traces, we make use of the 
Schwinger-DeWitt (SDW) technique, to get

\begin{equation}
{\rm Tr} \ln \hat{\cal F} = \frac{i}{ (4 \pi)^{\frac{d}{2}} } 
\int_0^{\infty}
\frac{ds}{s^{\frac{d}{2}+1}} \int d^dx 
{\rm Tr} \sum_{n=0}^{\infty} (i s)^n \hat{a}_n(x) ,
\end{equation}
where the $\hat{a}_n(x)$'s are the coincidence limit of the SDW 
coefficients. The divergent part of the EA for any minimal operator
in $d=4$ dimensions is determined by the first three SDW coefficients.
The divergences coming from $\hat{a}_0$ and $\hat{a}_1$ can be absorbed into
a redefinition of the cosmological constant and the Newton constant. In what
follows it will be relevant the divergence coming from the second SDW 
coefficient. It reads

\begin{equation}
\hat{a}_2(x) = \frac{1}{180} (R_{\mu\nu\alpha\beta} R^{\mu\nu\alpha\beta} -
R_{\mu\nu} R^{\mu\nu} + \Box R) \hat{1} + \frac{1}{2} \hat{\cal Q}^2 
+ \frac{1}{12} \hat{\cal R}_{\mu\nu} \hat{\cal R}^{\mu\nu} + \frac{1}{6} 
\Box \hat{\cal Q} ,
\end{equation}
where $\hat{\cal R}_{\mu\nu}$ 
is the commutator of covariant derivatives. Inserting
the definition of the operators ${\hat Q}$ for the graviton and the
ghost parts into the formula for the second SDW coefficient, one can
extract the divergence coming from pure gravity and the corresponding one
due to the massive terms. The former one gives the well-known result \cite{THV}

\begin{eqnarray}
\Delta S_G^{\rm div}(\lambda=0) &=&
\frac{2}{(4-d) 96 \pi^2} \int d^4x \sqrt{-g}
\left[ 
\frac{53}{15} (R_{\mu\nu\rho\sigma} R^{\mu\nu\rho\sigma} - 
4 R_{\mu\nu} R^{\mu\nu} + R^2) + \right. \nonumber \\
&& \left. ~~~~~~~~~~~~~~~~~~~~~~~~~~~~~~~~~
\frac{21}{10} R_{\mu\nu} R^{\mu\nu} +
\frac{1}{20} R^2 
\right] ,
\end{eqnarray}
where the first term in brackets is the Gauss-Bonnet term, a topological
invariant in $d=4$ spacetime dimensions. The divergence due to the presence
of $M$ is read from its contribution to the second SDW coefficient, namely
$\frac{1}{6} \Box \hat{M} + \frac{1}{2} \hat{M}^2 + \hat{M} 
(\hat{P} + \frac{1}{6} R \hat{1})$. It has the following form

\begin{equation}
\Delta S_M^{\rm div}(\lambda=0) = \frac{2}{(4-d) 64 \pi^2} \int d^4x \sqrt{-g}
\left[
M_{\mu\nu\rho\sigma} M^{\mu\nu\rho\sigma} + 2  M_{\mu\nu\rho\sigma}  
\left( P^{\rho\sigma\mu\nu} + \frac{1}{6} R \delta^{\rho(\mu} 
\delta^{\sigma\nu)} \right) \right] .
\end{equation} 
where we have omitted the boundary term.

%%%%%%%%%%%%%%%%%%%%%%%%%%%%%%%%%%%%%%%%%%%%%%%%%%%%%%%%%%%%%%%%%%%%%%%%%

\section{DIVERGENCES FOR NON-MINIMAL GAUGES}

In this Appendix we sketch the calculation of the one loop EA and its
divergences for the $\lambda$-family of gauge fixing functions. As we
have already remarked in the text, when $\lambda \neq 0$ we have a 
non-minimal gauge. For these non-minimal gauges, a reduction method has
been developed in \cite{BV2} which generalizes the Schwinger-DeWitt expansion.
It also consists in a local expansion in the background fields, and it has 
been calculated up to second
order in the curvature tensors. The starting point is to note that, since the
theory as a whole is gauge independent on the mass shell, the difference of
the effective action in any gauge from that in a given minimal gauge is 
always proportional to the extremal, i.e. the lhs of the classical
field equation. With this idea in mind, that difference can be expressed in
terms of non minimal Green's functions for gravitons and ghosts, which are
expanded in terms of the background dimensionality. 

One special easy case of non-minimal gauge families is that when the 
gauge-breaking action differs from the minimal one only by an overall factor.
This is indeed the case for the $\lambda$-family, since 
 $\chi^{\mu}(\lambda) = \frac{1}{\sqrt{1+\lambda}} \chi^{\mu}(\lambda=0)$.
Following the methods of \cite{BV2}, the EA for any member of this family
of gauge fixing functions is

\begin{equation}
S_{{\rm eff}}(\lambda) = S_{{\rm eff}}(\lambda=0) + \frac{i}{2} \lambda 
\left[ 
{\rm Tr} V_{1\nu}^{~\mu}(\nabla) - {\rm Tr} V_{2\nu}^{~\mu}(\nabla)  
\right]
-\frac{i}{4} \lambda^2 {\rm Tr} [ V_{1\nu}^{~\mu}(\nabla)]^2
+ {\cal O} (({\cal E}^{\mu\nu})^2) ,
\end{equation}
where the extermal ${\cal E}^{\mu\nu}$ is given by

\begin{equation}
{\cal E}^{\mu\nu} = \frac{\delta(S_G+S_M)}{\delta g^{\mu\nu}} = 
-\frac{2}{\kappa^2} (R^{\mu\nu}-\frac{1}{2} R g^{\mu\nu})
+\frac{1}{2} T^{\mu\nu} ,
\end{equation}
and $V_{1\nu}^{~\mu}(\nabla)$ and $V_{2\nu}^{~\mu}(\nabla)$ are tensors
that are linear and quadratic in the extremal respectively.
Their action on a test function $\zeta^{\nu}$ is given by

\begin{eqnarray}
V_{1\nu}^{~\mu}(\nabla) \zeta^{\nu} &=& 
2 \kappa^2 \, Q^{\mu}_{\alpha} \,  \nabla_{\beta} 
\, \Gamma^{(\alpha}_{\rho\sigma}(\nabla) \, {\cal E}^{\rho\beta)} \,
Q^{\sigma}_{\nu} \, \zeta^{\nu} , \nonumber\\
V_{2\nu}^{~\mu}(\nabla) \zeta^{\nu} &=&
- \kappa^2 g^{\mu\omega} \, Q^{\gamma}_{\omega} 
\, {\cal E}^{(\alpha\rho} \, \Gamma^{\beta)}_{\rho\gamma}(\nabla) \,
G_{\alpha\beta,\varphi\theta}(\nabla)  \, 
\Gamma^{(\varphi}_{\delta\sigma}(\nabla) \, {\cal E}^{\theta)\delta} \,
Q^{\sigma}_{\nu} \, \zeta^{\nu} .
 \end{eqnarray}
In these expressions, $\Gamma^{\nu}_{\rho\sigma}(\nabla)=
\delta^{\nu}_{\rho} \nabla_{\rho} - 2 \delta^{\nu}_{\sigma} \nabla_{\rho}$, 
and $G_{\alpha\beta,\varphi\theta}(\nabla)$ and 
$Q^{\sigma}_{\mu}$ are the Green's functions for the
gauge field and ghost field respectively, evaluated for the DeWitt gauge,
\begin{eqnarray}
F^{\gamma\sigma,\alpha\beta}(\nabla|\lambda=0) \,
G_{\alpha\beta,\varphi\theta}(\nabla)&=& -
\delta^{\gamma\sigma}_{\varphi\theta} ,
\nonumber\\
(\Box \delta^{\mu}_{\alpha} + R^{\mu}_{\alpha}) Q^{\sigma}_{\mu} &=&
\delta^{\sigma}_{\alpha} .
\end{eqnarray}

We are interested just in the contribution to 
the EA that is linear in the extremal (see main text). Therefore we concentrate
ourselves on ${\rm Tr} V_{1\nu}^{~\mu}(\nabla)$, which is given by
\begin{equation}
{\rm Tr} V_{1\nu}^{~\mu}(\nabla) = 2 \kappa^2 \int d^4x 
\left[
R^{\alpha}_{\, \cdot \, \gamma\beta\sigma} {\cal E}^{\gamma\beta} -
{\cal E}^{\beta\gamma} \delta^{\alpha}_{\sigma} \nabla_{\beta} 
\nabla_{\gamma}
\right]  
(\Box \delta^{\sigma}_{\alpha} + R^{\sigma}_{\alpha})^{-2}
\delta(x,y) |_{y=x} .
\end{equation}
In order to calculate the divergent part of this expression we use the methods
explained in \cite{BV2}. It is worth recalling that we are working up to
quadratic order in curvatures, so that for the contribution of the first
term in brackets we can approximate 
 $(\Box \delta^{\sigma}_{\alpha} + R^{\sigma}_{\alpha})^{-2}$ by
 $\Box^{-2} \delta^{\sigma}_{\alpha}$. The two
divergences that appear are 
\begin{eqnarray}
\Box^{-2} \delta^{\sigma}_{\alpha} \delta(x,y) |^{{\rm div}}_{y=x} &=&
\frac{i}{8 \pi^2} \frac{1}{4-d} \sqrt{-g} , \\
\nabla_{\beta} \nabla_{\gamma}
(\Box \delta^{\sigma}_{\alpha} + R^{\sigma}_{\alpha})^{-2}
\delta(x,y)|^{{\rm div}}_{y=x} &=&
\frac{i}{8 \pi^2} \frac{1}{4-d} \sqrt{-g} 
\left[
\frac{1}{6} (R_{\beta\gamma}-\frac{1}{2} g_{\beta\gamma} R) 
\delta^{\sigma}_{\alpha} + \right. \nonumber \\ 
&& ~~~~~~~~~~~~~~~~~~~~~
\left. \frac{1}{2} R^{\sigma}_{\, \cdot \, \alpha\beta\gamma} - 
\frac{1}{2} g_{\beta\gamma} R^{\sigma}_{\alpha}
\right] .
\end{eqnarray}
Finally the total divergence reads

\begin{equation}
\left. {\rm Tr} V_{1\nu}^{~\mu}(\nabla) \right|^{{\rm div}} =
\frac{2 i}{4-d} \, \frac{\kappa^2}{24 \pi^2} \int d^4x \sqrt{-g}
\left[ -5 R_{\mu\nu} {\cal E}^{\mu\nu} +
\frac{5}{2} R g_{\mu\nu}  {\cal E}^{\mu\nu} \right] .
\end{equation}

%%%%%%%%%%%%%%%%%%%%%%%%%%%%%%%%%%%%%%%%%%%%%%%%%%%%%%%%%%%%%%%%%%%%%%%%%%%%
%
%		References
%
%%%%%%%%%%%%%%%%%%%%%%%%%%%%%%%%%%%%%%%%%%%%%%%%%%%%%%%%%%%%%%%%%%%%%%%%%%%%

%%%%%%%%%%%%%%%%%%%%%%%%%%%%%%%%%%%%%%%%%%%%%%%%%%%%%%%%%%%%%%%%%%%%%%%%%

\end{document}